\documentclass[pre,amsmath]{revtex4-2}
\usepackage[pdftex]{graphicx} 
\usepackage{color}
\usepackage{bm}
\usepackage{epsfig}
\usepackage{latexsym}
\usepackage{nameref,hyperref}
\usepackage{xcolor}

\begin{document}
\newcommand{\be}{\begin{equation}}
\newcommand{\ee}{\end{equation}}
\newcommand{\bea}{\begin{eqnarray}}
\newcommand{\eea}{\end{eqnarray}}
\title{Run-and-tumble chemotaxis using reinforcement learning }
\author{Ramesh Pramanik$^{1}$, Shradha Mishra$^2$ and Sakuntala Chatterjee$^1$}
\affiliation{(1) Physics of Complex Systems, S.N. Bose National Centre for Basic Sciences, Salt Lake Sector 3, Kolkata 700106, India \\
(2) Department of Physics, Indian Institute of Technology (BHU) Varanasi 221005, India}
\begin{abstract}
Bacterial cells use run-and-tumble motion to climb up attractant concentration gradient in their environment. By extending the uphill runs and shortening the downhill runs the cells migrate towards the higher attractant zones. Motivated by this, we formulate a reinforcement learning (RL) algorithm where an agent moves in one dimension in the presence of an attractant gradient. The agent can perform two actions: either persistent motion in the same direction or reversal of direction. We assign costs for these actions based on the recent history of the agent's trajectory. We ask the question: which RL strategy works best in different types of attractant environment. We quantify efficiency of the RL strategy by the ability of the agent (a) to localize in the favorable zones after large times, and (b) to learn about its complete environment. Depending on the attractant profile and the initial condition, we find an optimum balance is needed between exploration and exploitation to ensure the most efficient performance.
\end{abstract}
\maketitle

\section{Introduction}

Reinforcement learning (RL) is a branch of machine learning \cite{sutton2018reinforcement, panait2005cooperative, busoniu2008comprehensive, kubat2017introduction, Goldberg1988, watkins1992q}, where an agent makes a choice to execute suitable actions to maximize its reward, while learning from its past \cite{durve2020learning}. In previous studies  RL has been used in many areas, such as game theory \cite{NIPS2005_9752d873}, operations research, information theory \cite{kanai}, statistical physics \cite{durve2020learning}, etc. In recent years RL has been successfully used to understand and control different aspects of many equilibrium and nonequilibrium systems. Recently in \cite{huang2021predicting, sampat2022ordering} RL is used to predict the nucleation and phase transition in two-dimensional Ising model. The results obtained from RL were in good match with the predictions from classical nucleation theory and from exact results. The learning strategy were also used in nonequilibrium systems like directed percolation in one and two dimensions \cite{shen2021supervised} and RL proved to be very successful in predicting the critical points. In \cite{kubo2022efficient} an efficient RL strategy is used in active flow control and recently RL is also used to navigate micro-swimmers to swim in response to hydrodynamic flow, shear flow and shear gradient \cite{pre23}. It was found that swimmers use their instantaneous orientation direction to obtain the optimal policy to perform efficient swimming in different environments.

In this work, we use reinforcement learning to study a model that has been motivated by the phenomenon of bacterial chemotaxis \cite{eisenbachbook, adler1973chemotaxis, adler1973method}. Certain bacteria like {\sl E.coli, S.typhimurium, B.subtilis} are known to show chemotaxis where they can  move along a chemical gradient in their environment \cite{berg2008coli, galloway1980histidine, sidortsov2017role}. When these microorganisms experience concentration gradient of an attractant chemical in their surroundings, they show a tendency to migrate towards regions of higher attractant concentration \cite{colin2017emergent, tu2013quantitative, lan2016information, parkinson2015signaling}. This migration happens via run-and-tumble motion, which is characterized by persistent movement along a particular direction (run), punctuated by abrupt change of direction (tumble) \cite{berg1972chemotaxis, Dev2019rnt}. In a homogeneous attractant environment, after a large number of runs and tumbles the net displacement of the cell is zero. But in presence of an attractant concentration gradient, runs in the favorable direction are extended and those in the opposite direction are shortened, giving rise to a chemotactic drift \cite{de2004chemotaxis, chatterjee2011chemotaxis, vladimirov2010predicted, dev2018optimal, shobhan}.

Among all types of microorganisms which show chemotaxis, {\sl E.coli} is the most well-studied one \cite{block1982impulse, keymer2006chemosensing, shimizu2010modular, tu2008modeling, bi2018stimulus, sourjik2002receptor}. The signaling network inside an {\sl E.coli} cell senses the attractant concentration gradient in the extra-cellular environment and regulates the run-and-tumble motion of the cell \cite{alon1999robustness, barkai1997robustness, frank2016networked, meir2010precision}. The detailed structure of the network has been studied both experimentally and by using theoretical models \cite{bray1998receptor, liu2012molecular, hansen2008chemotaxis, sourjik2004functional, jiang2010quantitative, mandal2022effect}. The spatial variation of attractant chemical across the cell body is negligible due to small size of the cell \cite{li2004cellular}. Instead, {\sl E.coli} senses the spatial gradient by comparing the attractant level experienced in the recent past and distant past along its run-and-tumble trajectory \cite{berg1975transient, kafri2008steady, celani2010bacterial, dev2015search}. If in the recent past attractant level is higher than that in the distant past, it generally means the cell is running up the concentration gradient. In that case, the cell tends to run for longer. If on the other hand, recent attractant level is lower than what the cell has experienced earlier, it tends to tumble quickly and chooses a new random direction to run. This mechanism allows the cell to accumulate in the favorable region \cite{vladimirov2008dependence, matthaus2009coli, long2017feedback, sun2017macroscopic}. After a long enough time has passed, the cell density becomes large in the region with high concentration of attractant.

Motivated by this, in the present paper, we formulate a reinforcement learning (RL) strategy where an agent performs run-and-tumble motion in an environment with inhomogeneous concentration of attractant. For simplicity, we consider one spatial dimension here. The agent can either persist moving in the same direction, or can reverse its direction. We define a cost matrix which assigns a cost to each of these two actions, depending on the recent history of the agent's trajectory.  With a small probability $\epsilon$ the agent `explores' its surroundings by performing a random action (persist or reverse) irrespective of its cost, and with the remaining probability $(1-\epsilon)$, the agent `exploits' its previous learning experience to decide its next action. The agent uses `$Q$-learning' method to learn and optimise its action based on its experience. $Q$-learning employs a non-supervised learning method and it is a simple version of RL which is based on optimizing a value function with respect to a given environment \cite{watkins1992q}.

We are interested in the long time behavior of the agent. In particular, we measure the effectiveness of the RL algorithm by evaluating the performance of the agent at large times. We use different performance criteria like how strongly the agent is able to localize in the high attractant zones, or how quickly it is able to find the region where the attractant concentration is highest, etc. The efficiency of the RL strategy can also be measured from how well the agent has learnt about its environment and whether this learning is used in its long time behavior. If the agent is found to behave based on only partial knowledge of its environment even after a long time has passed, then the RL strategy is deemed rather inefficient.

We consider two different types of attractant concentration profiles. In one case all the concentration peaks are of the same height, and in the other case they have unequal heights. We use a sine function, and superposition of two sine functions to represent these two cases (also see Fig. \ref{fig:env}) but our conclusions do not depend on the specific functional forms used. We consider a uniform initial condition, where the initial position of the agent is equally likely to be anywhere in the system, and a non-uniform initial condition where the agent starts from the vicinity of one particular peak of the attractant profile. For an attractant profile whose peak heights are equal, a uniform initial condition gives rise to a long time performance that gets better with less exploration (smaller $\epsilon$) and higher learning rate. However, for a non-uniform initial condition, a competition between exploration and exploitation ensues, which yields optimum ranges for exploration parameter and learning rate at which the RL strategy is most successful. When the attractant profile consists of peaks of different heights, even a uniform initial condition leads to optimal values for exploration and learning parameters. For a non-uniform initial condition when the agent starts near the lower attractant peak, we find that even after a long time, the agent is not able to preferentially localize itself near the higher attractant peak, unless exploration and learning parameters are chosen from an optimal range. For optimal choices of these two parameters, the long time behavior of the agent takes into account its full environment. Outside this optimal range, RL algorithm works inefficiently and the performance of the agent dips. We also measure the mean first passage time of the agent starting from the lower attractant peak to the higher peak. The mean first passage time shows a minimum for specific values of exploration and learning parameters. For these values the agent acts as the most efficient searcher and finds the higher peak in the shortest possible time. 

In Sec. \ref{sec:model} we describe the model in details and explain the RL algorithm. In Sec. \ref{sec:sin} we present our results for the sine wave attractant profile and in Sec. \ref{sec:2sin} we take up superposition of two sine waves. We include some concluding remarks in Sec. \ref{sec:con}.

\section{Formulation of RL algorithm} \label{sec:model}

\begin{figure}[htbp!]
\centering
\includegraphics[scale=0.15]{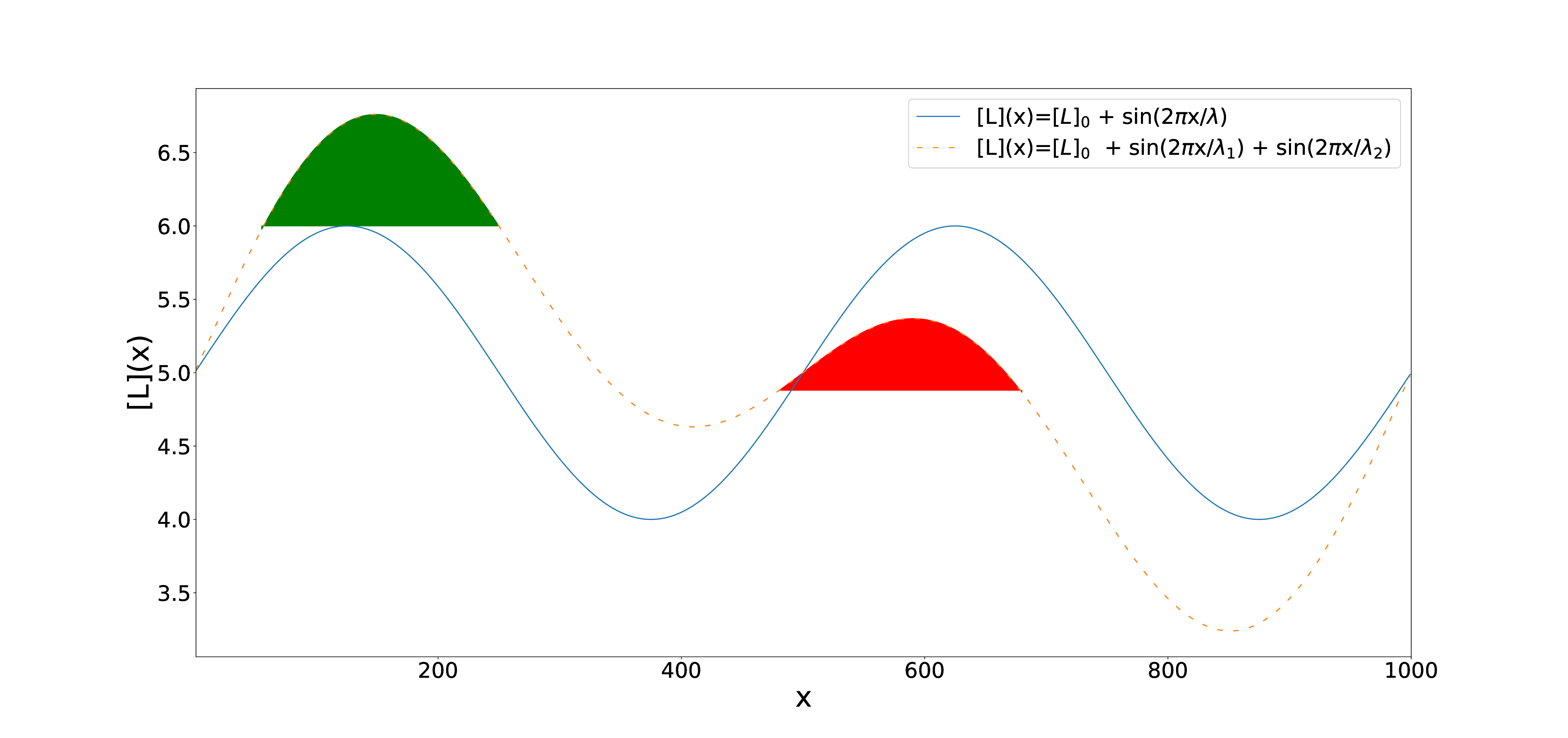}
\caption{Attractant concentration profile $[L](x)$. Solid line corresponds to $[L](x) = [L]_0+  \sin (2 \pi x / \lambda)$ and dashed line corresponds to $[L](x) = [L]_0+ \sin (2 \pi x / \lambda_1) + \sin (2 \pi x / \lambda_2)$. Here, $[L]_0$ represents the background concentration, which also makes sure $[L](x)$ never becomes negative. Unless mentioned other wise, we throughout use $[L]_0 = 5$, $\lambda = \lambda_1 = 500$ and $\lambda_2=1000$ and system size $L=1000$ with periodic boundary condition.}
\label{fig:env}
\end{figure}

We consider a one dimensional system of size $L$ with periodic boundary condition across which a spatially varying attractant concentration profile $[L](x)$ is set up. A successful RL strategy should allow the agent to efficiently localize near the maxima of $[L](x)$. Here, we mainly consider two scenarios: one in which $[L](x)$ has multiple maxima of the same height, and the other in which different maxima of $[L](x)$ have different heights. We choose simple functional forms, a sine wave and superposition of two sine waves to represent the above two scenarios, although our results and arguments for these two cases remain valid for any profile with equal and unequal peaks, respectively. In Fig. \ref{fig:env} we plot these two functions. For the sake of comparison we also consider a flat attractant profile, where $[L](x)$ is constant for all $x$. 

A particle (or agent) can be in two different states, `high' or `low', depending on the local attractant level. If the  attractant concentration at the location of the agent is equal to or higher than the background level $[L]_0$, the agent is assigned a state `high', and if the local concentration is lower than $[L]_0$, the state is `low'. The agent moves on the one dimensional system, either leftward or rightward, with fixed velocity $v$. In a time-step $dt$ its displacement is $vdt$. This sets the lattice spacing in our system and we use $|v|dt$ as the unit of length. Time is measured in units of $dt$. During a time-step the agent can perform two actions: `persist' or `reverse', {\sl i.e.} it can either retain its current run direction, or it can start running in the opposite direction. Below we describe how to calculate the cost for each of these actions.

In line with the strategy employed by the bacteria during chemotaxis, in our model the agent calculates the cost for each action based on its recent history. By comparing the attractant concentration experienced in the recent past and distant past, the agent assigns a cost for persisting in the current direction and reversing its direction. Define $x_1 \equiv x(t-\Delta_1)$ as the position of the agent at a time $\Delta_1$ back in the past, and similarly, $x_2 \equiv x(t-\Delta_2)$ with $\Delta_2 > \Delta_1$. If $[L](x_1) \geq [L](x_2)$ then we assign a cost $0$ for persisting and cost $1$ for reversing. On the other hand, if $[L](x_1) < [L](x_2)$, the cost for persisting is $1$ and cost for reversing is $0$. This method of assigning cost is consistent with the fact that in bacterial chemotaxis the runs in favorable directions are extended and those in unfavorable directions are shortened. Unless mentioned otherwise, throughout this work we use $\Delta_1 =1$ and $\Delta_2 =2$, {\sl i.e.} the agent compares the attractant encountered between last two time-steps (see Sec. \ref{sec:con} also for a detailed discussion on this choice). Our method of cost calculation also shows that the background concentration $[L]_0$ does not affect anything, since we are only interested in the difference between $[L](x_1)$ and $[L](x_2)$. In Table \ref{tab} we show the relevant parameters.

\begin{table}[htbp!] 
\caption{Table for model parameters}
\begin{tabular}{|c | c |}
 \hline
 Symbol & Value \\
 \hline
 $L$ & $1000$ \\
 
$\Delta_1 $  & $1$  \\
 
$\Delta_2 $  & $2$ \\

$p_0$ &  $0.9$  \\

$\alpha$  & Range $[0, 0.5]$\\  

$\epsilon$  & Range  $[0,1]$  \\

$\lambda$  & $500$   \\

$\lambda_1$ & $500$ \\

$\lambda_2$ & $1000$ \\
 
\hline
\end{tabular}
\label{tab}
\end{table}

In any run-tumble motion, the probability to tumble in a given time-step remains much less than the probability to keep running. When these two probabilities become equal, it ceases to be a run-tumble motion and becomes ordinary diffusion instead. Consistent with this, we assume that at every time-step the agent persists in the same direction with a probability $p_0 \gg 0.5$ and with the remaining probability $(1-p_0)$ it employs the RL algorithm to decide its course of action.

In RL algorithm with $Q$-learning, the information about previous learning experience is encoded in the $Q$-matrix whose row and column indices correspond to different states and actions, respectively. For our model with two possible states (high/low) and two possible actions (persist/reverse), the $Q$-matrix is $2 \times 2$ dimensional, with rows (columns) representing the states (actions). $S=1(2)$ corresponds to high (low) state and $A=1(2)$ represent persist (reverse) action. At each time-step the agent at state $S$ estimates the cost of both actions $c[S,A]$ and estimates the $Q$ matrix elements corresponding to the row $S$ according to the formula 
\be
Q_{t+1}[S,A] \leftarrow Q_t [S,A]+\alpha (c[S,A] -Q_t [S,A]) \label{eq:Q}
\ee
with $A=1,2$, and the parameter $\alpha$ is called the learning rate. The agent follows the $\epsilon$-greedy algorithm, where with probability $(1-\epsilon)$ the agent `exploits' the previous learning experience and selects the action for which the $Q$-matrix element is minimum. With probability $\epsilon$ the agent `explores' its environment by choosing an action at random, irrespective of its cost. If at time $t$ the agent persists in the same direction, then $Q_{t+1}(S,1)$ is updated according to Eq. \ref{eq:Q} and $Q_{t+1}(S,2)$ value reverts back to $Q_t (S,2)$, since that action was not executed. Similarly, if the agent reverses its direction at time $t$, then $Q_{t+1}(S,2)$ is updated and $Q_{t+1}(S,1)$ remains same as $Q_t (S,1)$. We start with the initial condition that at $t=0$ all elements of the $Q$-matrix are zero. Below we illustrate the algorithm with a specific example.

$\bullet$ Suppose at time $t$ the agent is found at position $x(t)$ with velocity $v$ which can be $\pm 1$. With probability $p_0$ the agent persists in the same direction, and $x(t+1)=x(t)+v$. No cost calculation is done in this case. 

$\bullet$ With probability $(1-p_0)$ the agent performs exploitation or exploration. Let us consider the specific example, where the agent is in the high state, $S=1$ and that it had been  going uphill for some time in the past, such that $[L](x_1)> [L](x_2)$. Then the cost $c[1,1]$ for persisting is $0$ and the cost for reversal $c[1,2]=1$. The agent computes $Q_{t+1}(1,1)$ and $Q_{t+1}(1,2)$ according to Eq. \ref{eq:Q}.

$\bullet$ With probability $(1-\epsilon)$ the agent performs exploitation. If $Q_{t+1}(1,1) \le Q_{t+1}(1,2)$, the agent persists in the same direction with probability $(1-p_0)$. If $Q_{t+1}(1,1) > Q_{t+1}(1,2)$, then the agent reverses its direction with probability $(1-p_0)$.

$\bullet$ With probability $\epsilon$ the agent performs exploration. In this case, it chooses one of the two actions randomly with probability $\dfrac{1-p_0}{2}$ for each.

$\bullet$ If persistence is chosen (either in exploitation or in exploration), the position is updated as $x(t+1)=x(t)+v$, and $Q_{t+1}(1,1)$ is updated, while $Q_{t+1}(1,2)$ is reassigned its old value $Q_t (1,2)$. If reversal is chosen, then $x(t+1)=x(t)-v$, and $Q_{t+1}[1,2]$ is updated but $Q_{t+1}[1,1]$ goes back to $Q_{t}[1,1]$. 

In a similar manner, the steps for other cases, like $[L](x_1) \leq [L](x_2)$ or $S=2$ can be outlined.

To test the above model, we consider the simple case of a spatially homogeneous attractant environment. According to our definition, the system can only be in $S=1$ in this case, and can never visit $S=2$. Since $[L](x_1) = [L](x_2)$ at all times, persistence (reversal) cost remains $0(1)$ always. If we start with the initial condition that all $Q$-matrix elements are zero, then they remain zero at all later times. For any other choice of initial $Q$, it is easy to see that $Q_t[1,1]$ and $Q_t[1,2]$ vanish for large $t$, while $Q_t[2,1]$ and $Q_t[2,2]$ never change with time, since that state is never visited.

\begin{figure}[htbp!]
\centering
\includegraphics[scale=0.65]{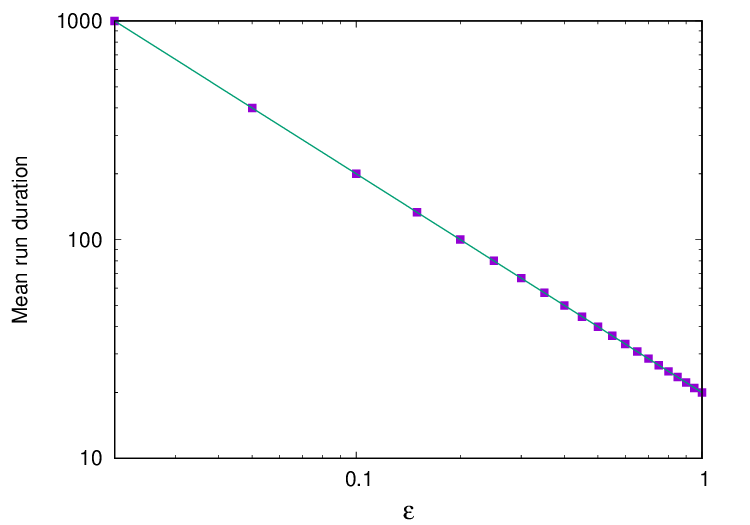}
\caption{The average run duration in a homogeneous medium is inversely proportional to $\epsilon$. The discrete points show simulation data, and the continuous line shows analytical expression. Here we have used $p_0=0.9$.} \label{fig:homo}
\end{figure}

From our algorithm described above, it follows that during exploitation the agent can never tumble in this case. Velocity reversal is possible if and only if the agent chooses reversal during exploration. The tumbling probability then becomes $\dfrac{\epsilon (1-p_0)}{2}$. The average number of steps between two successive tumbling events, or the mean run duration is then $\dfrac{2}{\epsilon(1-p_0)}$. Therefore, in a homogeneous environment we retrieve the simple run-tumble motion with constant tumbling rate, as expected. We explicitly compare our calculation with simulation results, in Fig. \ref{fig:homo}.

\section{Performance of RL agent in sine wave attractant environment}
\label{sec:sin}

\begin{figure}[htbp!]
\includegraphics[scale=0.7]{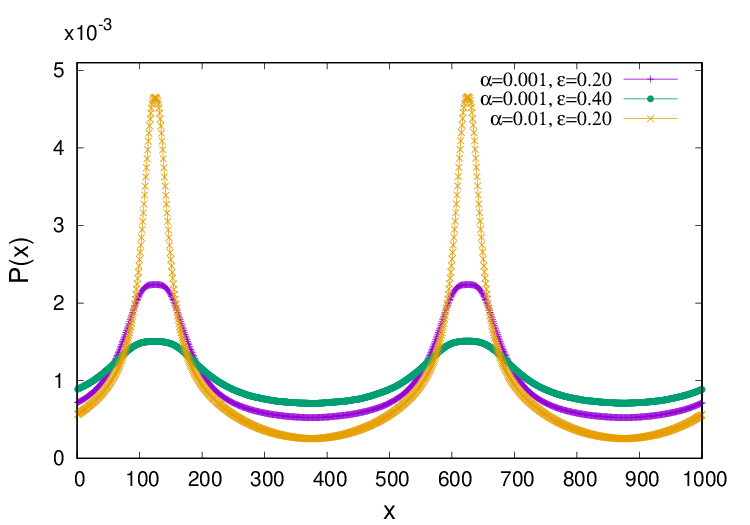}
\caption{The position distribution of the agent after a long time in a  sine wave attractant profile. The agent localizes more strongly near the two attractant peaks as $\epsilon$ decreases and/or $\alpha$ increases. Here we have used uniform initial condition and $\Delta{_1}=1$, $\Delta{_2}=2$, $p_0=0.90$.} \label{fig:sinpx}
\end{figure}

In this section, we present our results for the case when the attractant concentration shows a sinusoidal variation in space, as shown in Fig. \ref{fig:env} (solid line). In Fig. \ref{fig:sinpx} we plot the position distribution $P(x)$ of the agent in the long time limit. We find that RL algorithm can successfully generate chemotactic behavior in this system, {\sl i.e.} the particle is more likely to be found near the maxima of the sine curve.  However, depending on the exploration parameter and learning rate, interesting variations are observed. 

\subsection{Shortest run for a specific exploration parameter}

\begin{figure}[hbt!]
\centering
\includegraphics[scale=0.7]{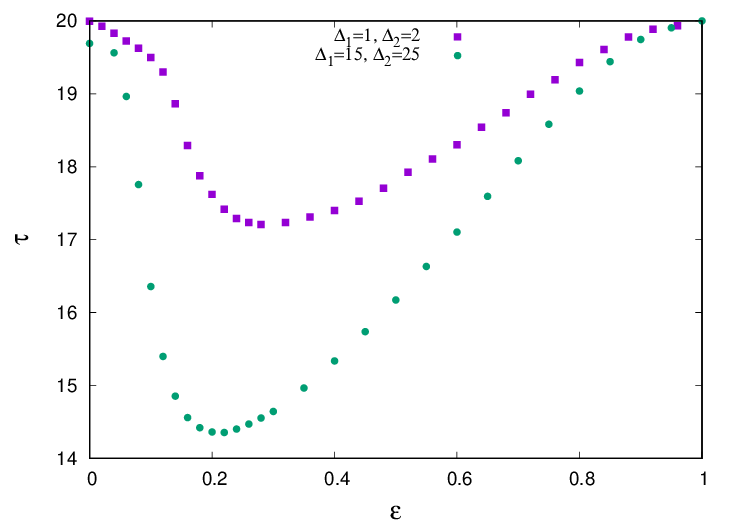}
\caption{Mean run duration $\tau$ shows a minimum with $\epsilon$. These data are for sine wave attractant profile with $p_0=0.90$ and $\alpha=0.001$.}
\label{fig:tau_ep}
\end{figure}

In Fig. \ref{fig:tau_ep} we plot the mean run duration $\tau$ as a function of $\epsilon$ and find a minimum. Our data also show that $\tau$ for $\epsilon=0$ and $1$ have similar values, particularly when $\Delta_1$ and $\Delta_2$ are small (purple squares). For $\epsilon=1$ the agent only explores and never exploits. At every time-step the agent persists in the same direction with probability $\dfrac{1+p_0}{2}$ and reverses its direction with probability $\dfrac{1-p_0}{2}$, irrespective of the local attractant environment. In this case its behavior is same as in the homogeneous case. $\tau$ therefore has the value $\dfrac{2}{1-p_0}$. For our choice of $p_0=0.9$ we find very good agreement with the simulation data.

The $\epsilon=0$ limit on the other hand, means that the agent never explores and only exploits at all times. When both $\Delta_1$ and $\Delta_2$ are smaller than $\tau$, then during an uphill run, reversal has cost $1$ and persistence has cost $0$. The agent in this case persists with probability $1$ until it reaches the point where attractant concentration is maximum. Beyond this point, the run continues as downhill run. Although the cost for persisting now becomes $1$ and reversal has no cost, the agent can still persist with probability $(1-p_0)$. The run terminates when the agent has crossed an average number of $\dfrac{1}{1-p_0}$ steps after crossing the peak. A new run starts in the opposite direction, this time going uphill with zero tumbling probability, and after crossing the attractant peak, the run again terminates after an average duration of $\dfrac{1}{1-p_0}$. This means in the steady state for $\epsilon =0$ the agent hovers around the peak of the attractant concentration profile and the mean separation between two successive tumbles, {\sl i.e.}  $\tau= \dfrac{2}{1-p_0}$, same as we found for $\epsilon=1$. This is consistent with our data in Fig. \ref{fig:tau_ep} (purple squares). 

Note that equal values for $\epsilon=0$ and $1$ means there has to be some maximum or minimum in between. It is possible to argue that $\tau$ must decrease with $\epsilon$ when $\epsilon$ is small but non-zero. In this case, because of exploration, the agent has a finite tumbling probability $\dfrac{(1-p_0)\epsilon}{2}$ during an uphill run. During a downhill run the tumbling probability is $(1-p_0)(1-\dfrac{\epsilon}{2})$. For a simple run-and-tumble motion with these tumbling rates in two directions, one can easily calculate the mean run duration and show that it decreases with $\epsilon$. Although in our present system the dynamics involves additional steps like $Q$ matrix calculation, etc., the qualitative trend for small $\epsilon$ is still captured by the above simplified description. This explains the minimum of $\tau$ as a function of $\epsilon$. In this argument, we assumed small values of $\Delta_1$ and $\Delta_2$, such that most of the time the comparison between recent and distant past is performed within the same run. When at least one of $\Delta_1$ and $\Delta_2$ becomes comparable to $\tau$, above argument does not work very well. In that case we find (Fig. \ref{fig:tau_ep} green circles) some deviation of the observed run duration from the analytical prediction $\dfrac{2}{1-p_0}$.

\subsection{Exploration-exploitation competition affects performance }

To quantitatively evaluate the performance of the agent, we define the following quantity, known as `uptake' \cite{celani2010bacterial, dev2018optimal},
\be 
\langle C \rangle = \int_0 ^L dx P(x) \{[L](x) -[L]_0\}
\label{eq:up}
\ee
Large uptake means in the long time limit the agent is more likely to be found in regions with high concentration of attractant, which indicates good performance. The maximum possible value of uptake is reached when the agent is localized at the peak of the attractant profile with all probability. It follows from Eq. \ref{eq:up} that the uptake in this limit is given by $[L](x)|_{max} - [L]_0$. In the case of a sine wave attractant profile which we consider here, largest possible uptake is $1$. Similarly, the lowest uptake is obtained when the agent localizes at the minimum of $[L](x)$ with all probability. In this case, the lowest uptake is $-1$. A negative uptake generally indicates a chemo-repellent environment ({\sl e.g.} some toxic chemical) from  which the agent prefers to stay away. When the uptake is zero, it means the agent is completely insensitive to the concentration gradient present in its environment and is equally likely to be found anywhere in the system.

\begin{figure}[htbp!]
\includegraphics[scale=0.6]{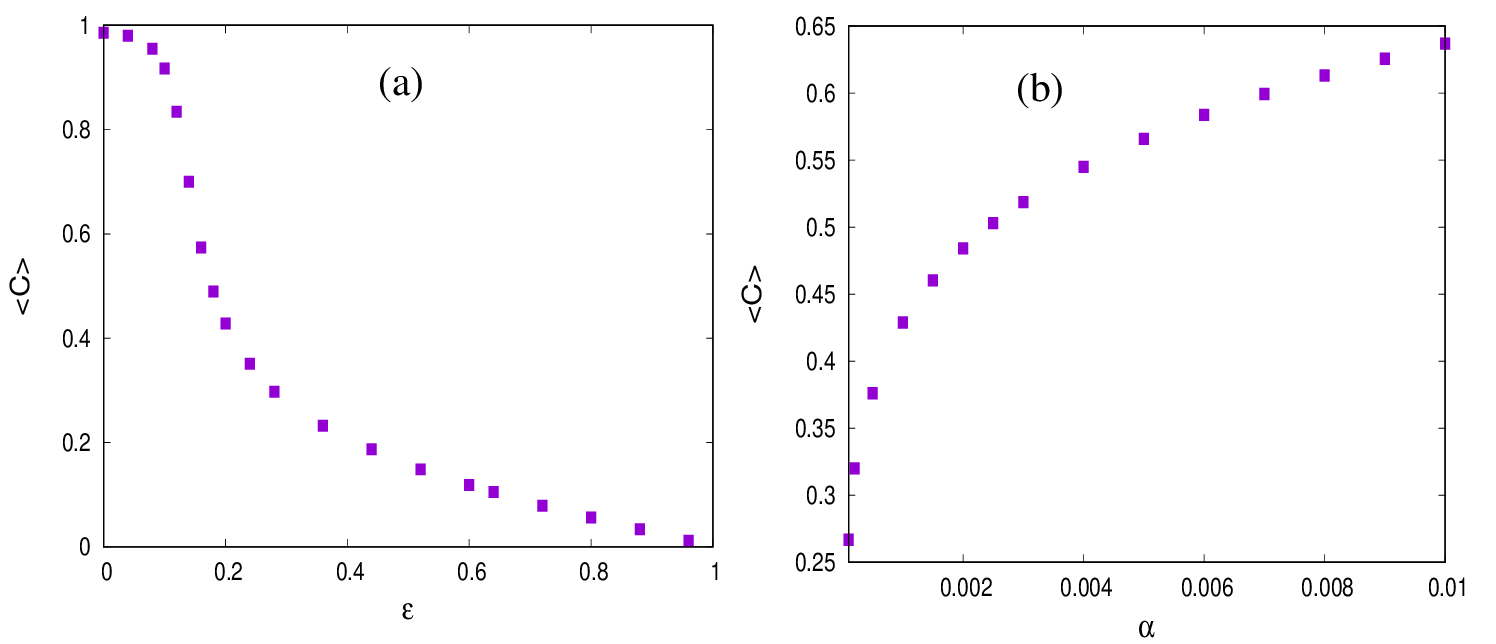}
\caption{Uptake $\langle C \rangle$ decreases with $\epsilon$ and increases with $\alpha$. In panel (a) we have used $\alpha=0.001$ and in (b) we have $\epsilon=0.20$. All other simulation parameters are as in Fig. \ref{fig:sinpx}. } 
\label{fig:epalp}
\end{figure}

In Fig. \ref{fig:epalp}(a) and (b) we plot the variation of uptake $\langle C \rangle$ as a function of the exploration parameter $\epsilon$ and learning parameter $\alpha$, respectively. Here, we have considered a uniform initial condition, {\sl i.e.} the initial position of the agent is chosen from a uniform distribution. We find that for a fixed $\alpha$ ($\epsilon$) uptake decreases (increases) with $\epsilon$ ($\alpha$). These data are consistent with our plots for $P(x)$ shown in Fig. \ref{fig:sinpx}. Both the Figs. \ref{fig:sinpx}, \ref{fig:epalp} show that increasing the learning rate $\alpha$ and/or decreasing the exploration parameter $\epsilon$ helps the agent to strongly localize near the peaks of $[L](x)$. However, we show below that for a different choice of initial condition, conclusions can be different.

Note that larger $\alpha$ or smaller $\epsilon$ inhibit the agent from sampling the whole environment. The agent remains confined near the peak of $[L](x)$. This can significantly affect its performance if the initial position of the agent is in the vicinity of one particular peak of $[L](x)$ (instead of being uniformly distributed over the whole system). In that case, the agent is not able to visit the other peaks of $[L](x)$ when $\alpha$ is too large or $\epsilon$ is too small. This can have a potentially adverse effect on the agent's performance when $[L](x)$ has peaks of different heights. The agent may get stuck near a local maximum and may not be able to reach the highest peak of $[L](x)$ even after a long time has elapsed. We explicitly consider such a case in Sec. \ref{sec:2sin}. For our present choice of a sinusoidal form of $[L](x)$ all the peaks are of same height. Therefore, even if the agent gets stuck near one peak, it will not affect a quantity like uptake. 
\begin{figure}[htbp!]
\includegraphics[scale=0.7]{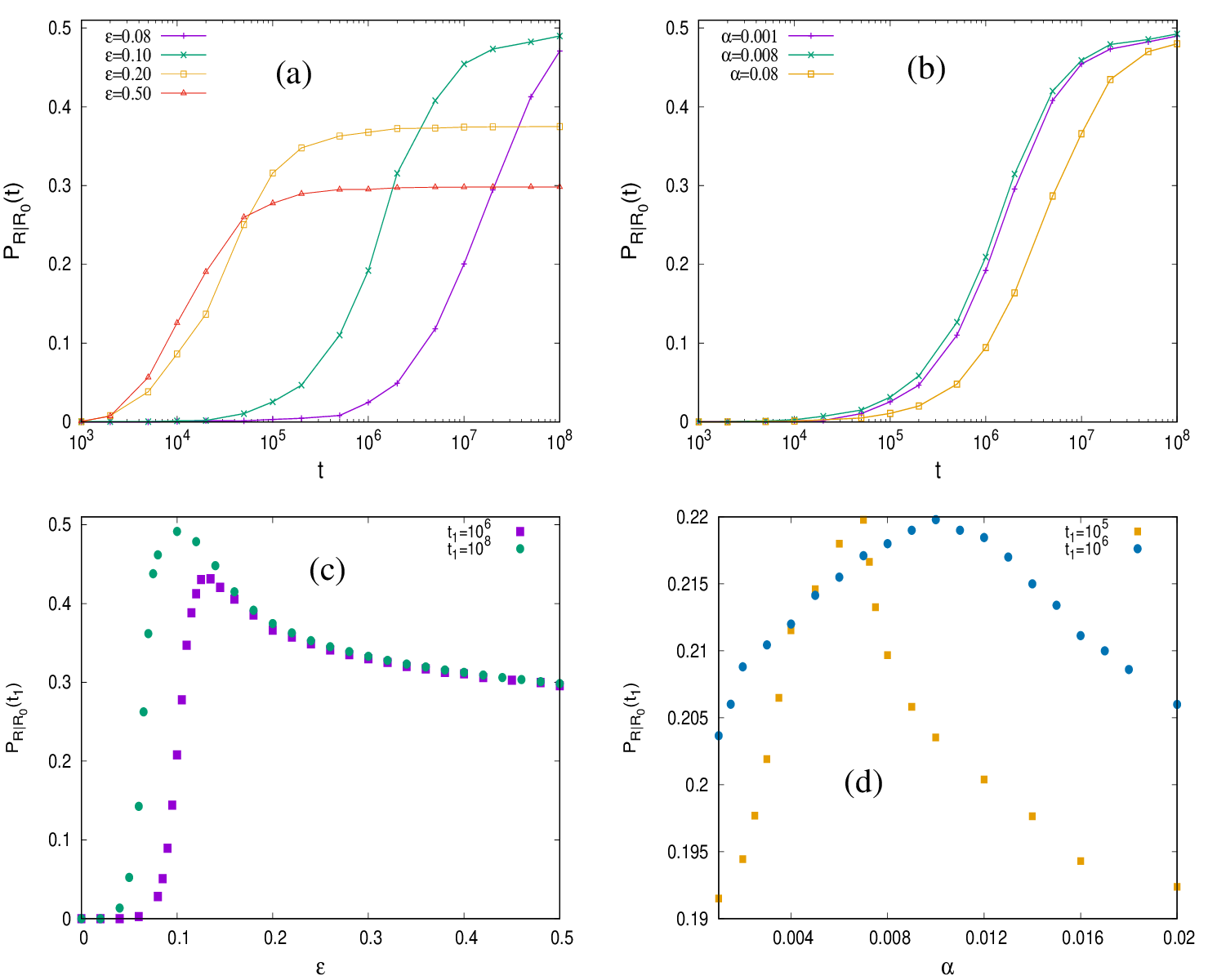}
\caption{Starting from the neighborhood of one attractant peak, the probability to find the agent near the other peak at large times. (a) ${\cal P}_{R|R_0} (t)$ increases with $t$ and saturates for large $\epsilon$ but as $\epsilon$ becomes small, due to trapping effect ${\cal P}_{R|R_0} (t)$ increases very slowly. Here, $\alpha$ is held fixed at $0.001$ (b) Similar effect observed as $\epsilon$ is held fixed at $0.1$ and $\alpha$ is varied. But $\alpha$-dependence is relatively weaker. (c) ${\cal P}_{R|R_0} (t=t_1)$ after a large time $t_1$ shows a peak as a function of $\epsilon$. Peak shifts to smaller $\epsilon$ for larger $t_1$. (d) ${\cal P}_{R|R_0} (t=t_1)$ also shows a peak with $\alpha$. For larger $t_1$ peak position shifts towards smaller $\alpha$ values. We have rescaled the data for $t_1=10^5$ by a constant factor $6.77$ in order to show both the plots in the same scale. In (a) and (c) we have used $\alpha = 0.001$ and in (b) and (d) $\epsilon = 0.1$. Other simulation parameters are as in Fig. \ref{fig:sinpx}.} \label{fig:p2p}
\end{figure}

To capture the trapping effect, therefore, we define another performance criterion, which measures if all the favorable regions in the environment has been sampled by the agent. Starting from the vicinity of one peak of $[L](x)$, we measure the probability to find the agent in the vicinity of another peak, as a function of time. More specifically, we choose the initial condition where the agent is equally likely to be located in a region $R_0$ with initial position lying in the range $0 < x < \dfrac{\lambda}{2} $. As a function of time we measure the probability to find the agent in the region $R$ which is in the neighborhood of the next peak with $\lambda < x < \dfrac{3 \lambda}{2}$. We denote this probability as ${\cal P}_{R|R_0}(t) $, which is expected to grow with time $t$ and then saturate as steady state is reached. However, as seen in our data in Figs. \ref{fig:p2p}(a) and (b) the saturation can be logarithmically slow with time, specially when $\epsilon$ ($\alpha$) is small (large). It is expected, since going from one peak to another, the agent needs to cross a region where $[L](x)$ is quite small. This means the agent has to execute long downhill runs which get increasingly difficult for small (large) $\epsilon$ ($\alpha$). For a logarithmically slow relaxation, it is often not feasible to evaluate the performance of the agent based on its steady state properties. Rather, the behavior of ${\cal P}_{R|R_0}(t) $ for large times (but not large enough to reach steady state) can be studied. If it is found that for a certain large value of $t=t_1$, the probability ${\cal P}_{R|R_0}(t_1) $ has low value, then that means the agent has not yet learned about its full environment, although a long time has passed. This indicates inefficient learning. On the other hand, a large value of ${\cal P}_{R|R_0}(t_1) $ means that the agent managed to learn about its environment and  determined its preference for a particular location based on this knowledge. The RL strategy in this case is working well.

In Fig. \ref{fig:p2p}(a) we plot $ {\cal P}_{R|R_0} (t) $ vs $t$ for different $\epsilon$ values and a fixed $\alpha$. As discussed in the previous paragraph, we find for small $\epsilon$ the probability increases almost logarithmically slowly with time but reaches high value eventually, whereas for larger $\epsilon$ the probability quickly saturates but the saturation value is lower. In Fig. \ref{fig:p2p}(c) we plot $ {\cal P}_{R|R_0} (t=t_1) $ as a function of $\epsilon$ for fixed $\alpha$ and show that the probability has a maximum value at a specific $\epsilon$. In other words, the performance measured after a certain large time $t_1$ is best at a specific $\epsilon$ value. As $t_1$ increases, the best performance is obtained at lower $\epsilon$. This trend is consistent with our expectation that eventually, for $t_1 \to \infty$ $ {\cal P}_{R|R_0} (t=t_1) $ will not depend on $t_1$ anymore and will saturate with time, with the highest saturation value obtained for $\epsilon =0$.

\begin{figure}[htbp!]
\includegraphics[scale=0.6]{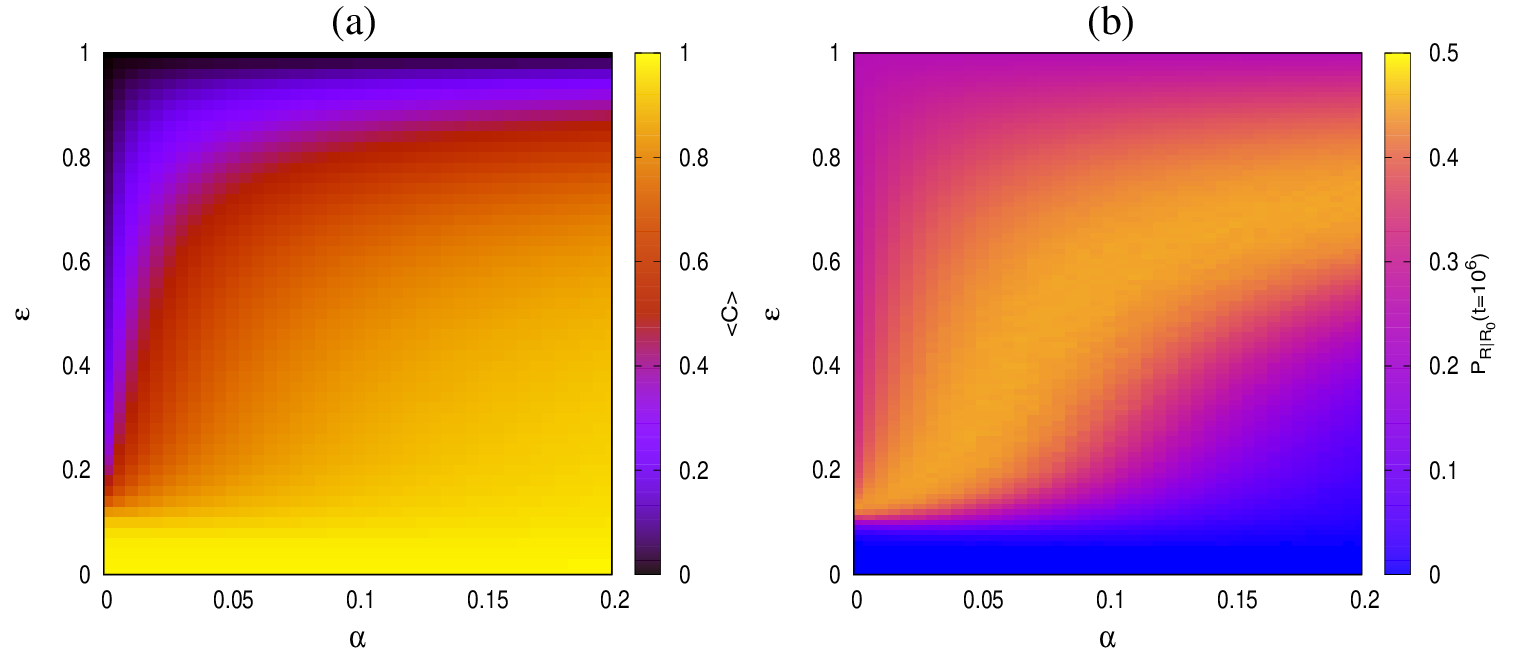}
\caption{Heat map of uptake $\langle C \rangle$ and $ {\cal P}_{R|R_0} (t=t_1)$ in the $\epsilon$-$\alpha$ plane. Here we have used $t_1=10^6$. While uptake is highest in the small-$\epsilon$ and large-$\alpha$ corner, $ {\cal P}_{R|R_0} (t=t_1)$ reaches its maximum value for an optimal $\epsilon$-$\alpha$ zone. Other simulation parameters are as in Fig. \ref{fig:sinpx}.} \label{fig:heat_sin}
\end{figure}

We find a similar qualitative behavior when we keep $\epsilon$ fixed and measure $ {\cal P}_{R|R_0} (t) $ for different values of $\alpha$. For small $\alpha$ the probability saturates faster but the saturation value is low. When $\alpha$ is large, $ {\cal P}_{R|R_0} (t) $ increases logarithmically slowly and ultimately saturates to a large value for very large times. When we measure $ {\cal P}_{R|R_0} (t=t_1) $ for a large $t_1$, we find a peak at a specific $\alpha$, as shown in Fig. \ref{fig:p2p}(d). However, the $\alpha$ dependence is rather weak, unlike the pronounced variation against $\epsilon$ shown by $ {\cal P}_{R|R_0} (t) $. In Fig. \ref{fig:heat_sin} we present the heat map of two performance criteria, uptake $\langle C \rangle $ (for uniform initial condition) and $ {\cal P}_{R|R_0} (t=t_1) $ (for initial position in the vicinity of one maximum of $[L](x)$) in the $\epsilon - \alpha$ plane. While uptake is maximum for smallest $\epsilon$ and largest $\alpha$ considered,  $ {\cal P}_{R|R_0} (t=t_1) $ shows a peak for intermediate values of $\epsilon$ and $\alpha$. 

\section{Attractant profile with different peak heights}
\label{sec:2sin}

\begin{figure}[htbp!]
\includegraphics[scale=0.8]{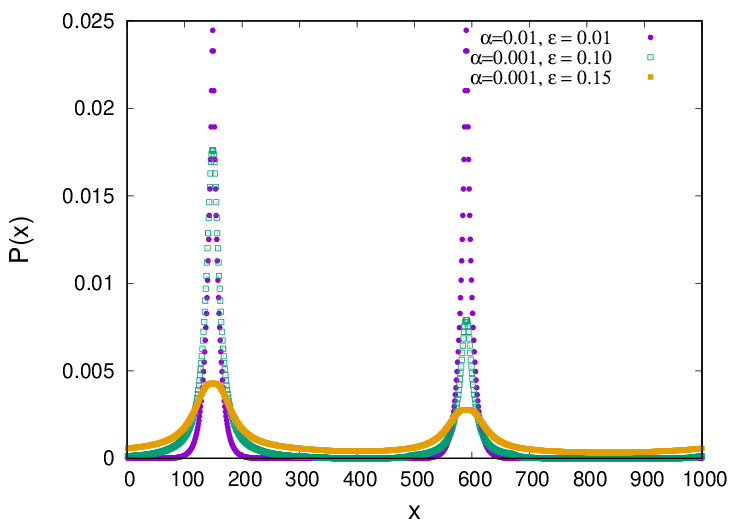}
\caption{Position distribution measured after time $t=10^6$ starting from a uniform initial condition. The attractant profile has peaks of unequal heights here. But for small $\epsilon$ and large $\alpha$ the agent shows almost equal affinity for both peaks, even after a long time. Here we have used $p_0 = 0.9$, $\Delta_1 =1$ and $\Delta_2 =2$.} \label{fig:px2sin}
\end{figure}

In this section, we consider an attractant concentration profile with two peaks of different heights, as shown by the dashed line in Fig. \ref{fig:env}. The functional form we consider here is given by $[L](x) = [L]_0 + \sin \dfrac{2 \pi x}{\lambda} + \sin \dfrac{\pi x}{\lambda}$.     We consider two different initial conditions, where the initial position of the agent is chosen from (a) a uniform distribution over the entire system, and (b) a uniform distribution in the vicinity of the lower peak (see Fig. \ref{fig:env}, region shaded by red). Like the previous section, we are interested in the long time performance of the agent for different choices of $\epsilon$ and $\alpha$. Due to the presence of high and low attractant peaks, the competition between exploration and exploitation is more clearly visible here. Even when we start from a uniform initial condition, for low $\epsilon$ and/or high $\alpha$, the agent may get trapped in the lower attractant peak. Uniform initial condition makes sure that the agent is equally likely to start from the vicinity of the lower and higher attractant peaks. For those trajectories which start near the lower (higher) peak, the agent quickly localizes itself at the lower (higher) peak, and remains trapped there for long time when $\epsilon$ ($\alpha$) is small (large). Thus even after a long time the agent is not able to learn about its complete environment and shows roughly the same preference for both peaks. Our data in Fig. \ref{fig:px2sin} show this trend clearly. 

\subsection{Performance peak for optimal $\epsilon$ and $\alpha$}
\begin{figure}[htbp!]
\includegraphics[scale=0.7]{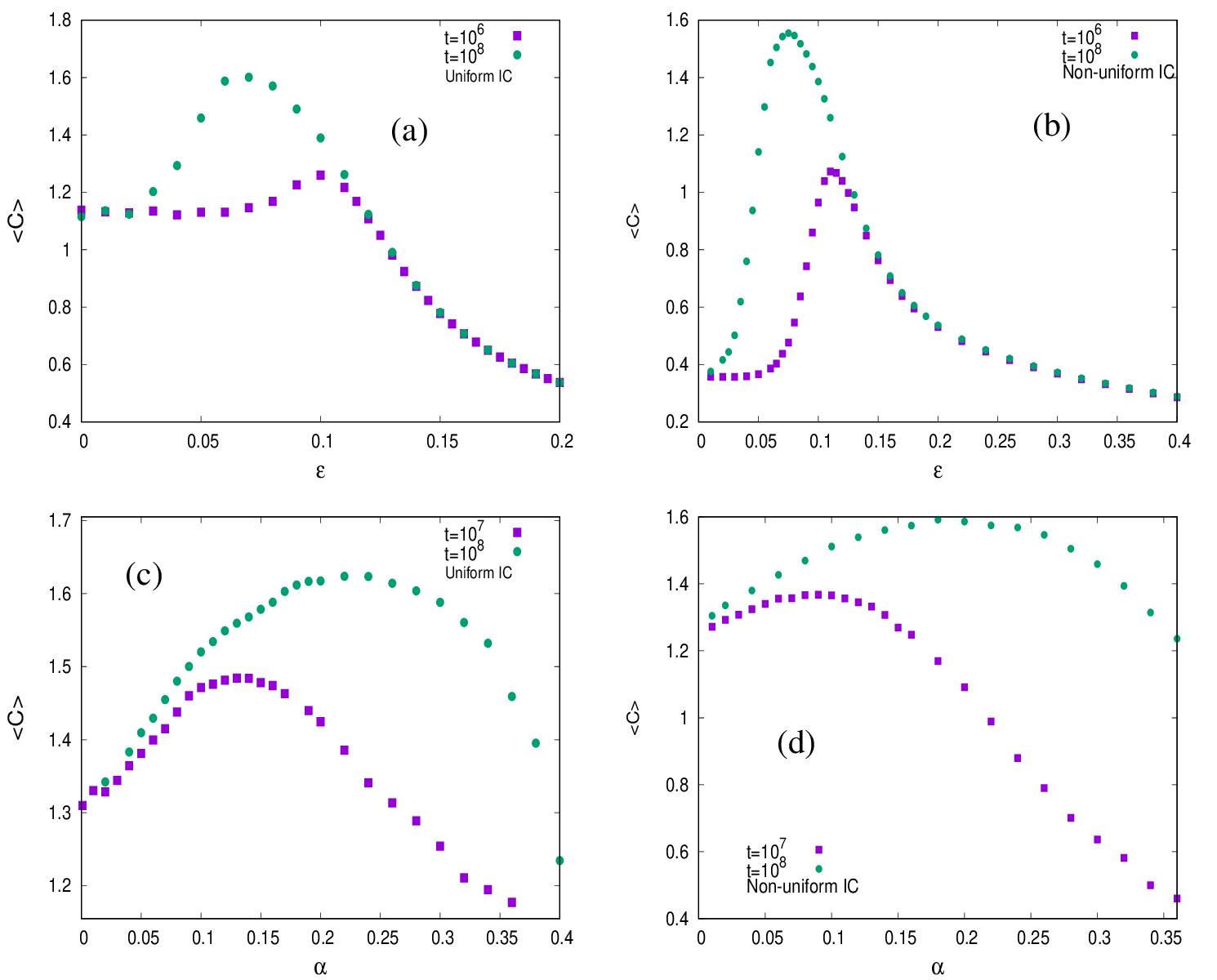}
\caption{Uptake $\langle C \rangle$ measured after a long time $t$ starting from uniform and non-uniform initial conditions.  As a function of $\epsilon$ and $\alpha$ the uptake shows a peak. For larger $t$ the peak shifts to smaller $\epsilon$ and larger $\alpha$. In (a) and (b) we have used $\alpha = 0.001$ and in (c), (d) we have $\epsilon = 0.1$. Other simulation parameters are as in Fig. \ref{fig:px2sin}. } \label{fig:2sincav}
\end{figure}

In Fig. \ref{fig:2sincav} we plot the value of uptake, measured after  long times, for different choices of $\epsilon$, $\alpha$ and starting with both types of initial conditions. In Fig. \ref{fig:2sincav}(a) uptake at two different times are plotted as a function of $\epsilon$, and for a fixed $\alpha$. These data are for uniform initial condition. As explained above, even after a long time, almost half of the population remains trapped near the lower peak of the attractant, when $\epsilon$ is small. This yields low value of uptake. On the other hand, when $\epsilon$ is too large, the exploitation loses out to exploration and the agent becomes less sensitive to the attractant gradient which decreases uptake. For intermediate $\epsilon$ uptake shows a peak. When a longer time has passed, a fraction of the population which was trapped near the lower attractant peak, gets a chance to find the higher peak. The uptake peak thus shifts to lower $\epsilon$ values. Fig. \ref{fig:2sincav}(b) shows the data for a non-uniform initial condition (where the agent starts near the lower attractant peak, shown by the region shaded in red in Fig. \ref{fig:env}). While the qualitative behavior remains similar, the uptake peaks at different times are much more pronounced here. This is not surprising since the trapping effect is now felt by the entire population. Figs. \ref{fig:2sincav}(c) and \ref{fig:2sincav}(d) show the plots for uptake measured at different times, as a function of $\alpha$ and for a fixed $\epsilon$, for uniform and non-uniform initial conditions, respectively. In this case, for large $\alpha$ effect of confinement is felt more strongly and uptake decreases. For very small $\alpha$, learning happens at a negligible rate, and the agent's preference for high attractant regions is low. Uptake is small in this limit too. For an intermediate $\alpha$ uptake has a peak and with time that peak shifts rightward towards larger $\alpha$ values.   

\begin{figure}[htbp!]
\includegraphics[scale=0.7]{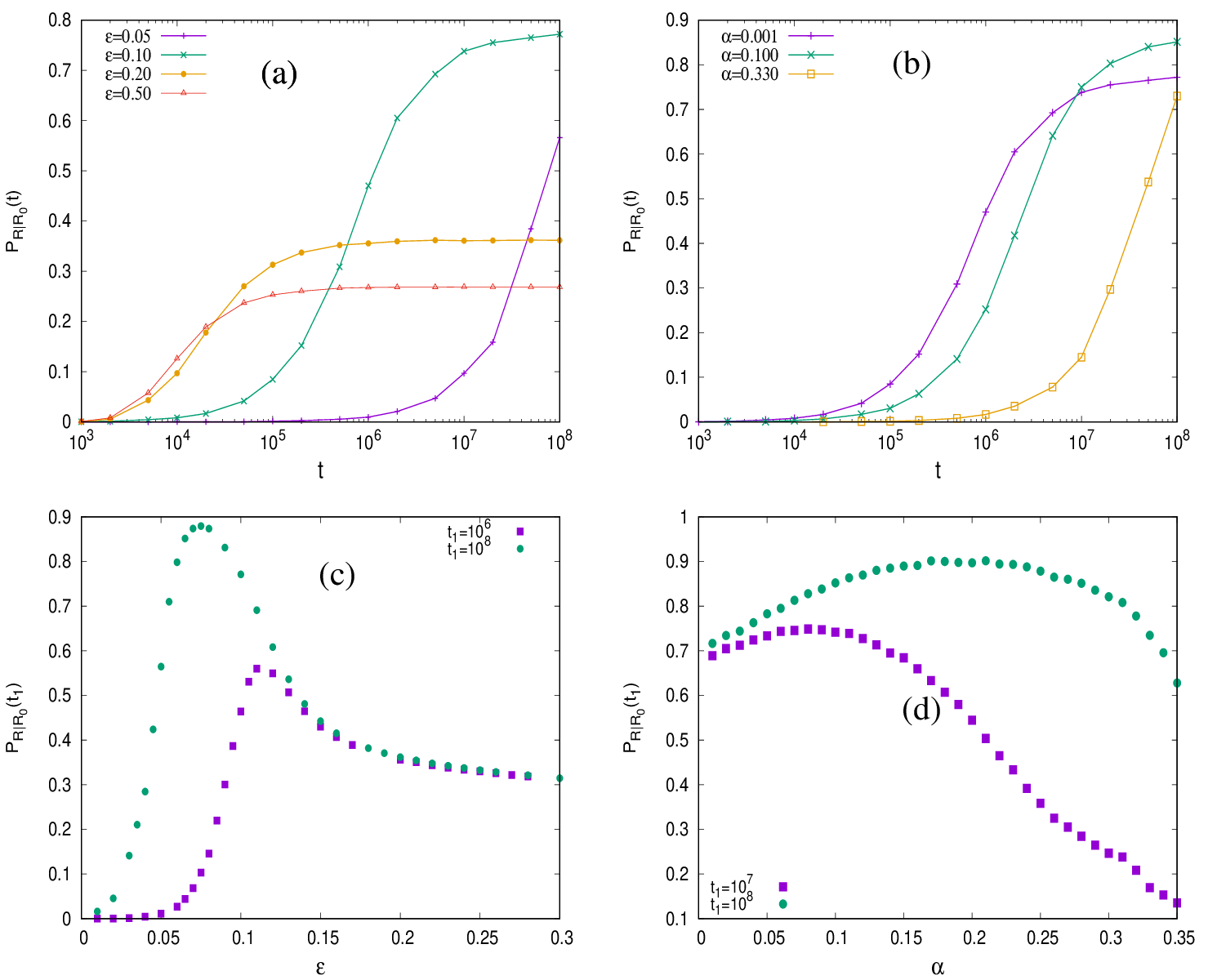}
\caption{Starting from a non-uniform initial condition, ${\cal P}_{R|R_0} (t)$ is measured for different $\epsilon$ and $\alpha$. (a) ${\cal P}_{R|R_0} (t)$  increases with $t$ and saturates quickly when $\epsilon$ is large. But as $\epsilon$ decreases the saturation comes at a later time and saturation value is also higher. Finally for very small $\epsilon$, ${\cal P}_{R|R_0} (t)$ increases logarithmically slowly with time. (b) Similar trend is observed as $\epsilon$ is held fixed and $\alpha$ is increased. (c) ${\cal P}_{R|R_0} (t=t_1)$ shows a peak as a function of $\epsilon$ and the peak position shifts leftward for larger $t_1$. (d) The peak of ${\cal P}_{R|R_0} (t=t_1)$ in $\alpha$ similarly shifts rightward towards larger $\alpha$ values when $t_1$ is increased. In (a) an (c) we have $\alpha=0.001$ and in (b), (d) we have $\epsilon = 0.1$. } \label{fig:2peak}
\end{figure}

We also measure the conditional probability ${\cal P}_{R|R_0} (t)$, as defined in Sec. \ref{sec:sin}, for different $\epsilon-\alpha$ value. Here, $R_0 $ represents the region shaded by red in Fig. \ref{fig:env} and the region $R$ is marked by green shade. The agent starts in the neighborhood of the lower peak of the attractant concentration and we measure the probability to find it in the vicinity of the higher peak as a function of time. The effect of trapping is much more pronounced in this case. While for large $\epsilon$ or small $\alpha$, Figs. \ref{fig:2peak}(a) and (c) show that  the probability ${\cal P}_{R|R_0} (t)$ increases with time and saturates quickly at a lower value, as $\epsilon$ decreases and/or $\alpha$ increases, ${\cal P}_{R|R_0} (t)$ shows a much slower growth with time, due to long trapping time of the agent at the lower peak. Fig. \ref{fig:2peak}(b) plots the value of ${\cal P}_{R|R_0} (t=t_1)$ at a fixed time $t_1$ as a function of $\epsilon$ and we find a peak. As $t_1$ increases, the peak shifts to lower $\epsilon$ values. Fig. \ref{fig:2peak}(d) similarly shows ${\cal P}_{R|R_0} (t=t_1)$ as a function of $\alpha$ and in this case peak shifts to larger $\alpha$ values. These trends are consistent with our data in Fig. \ref{fig:p2p}.

\begin{figure}[htbp!]
\includegraphics[scale=0.7]{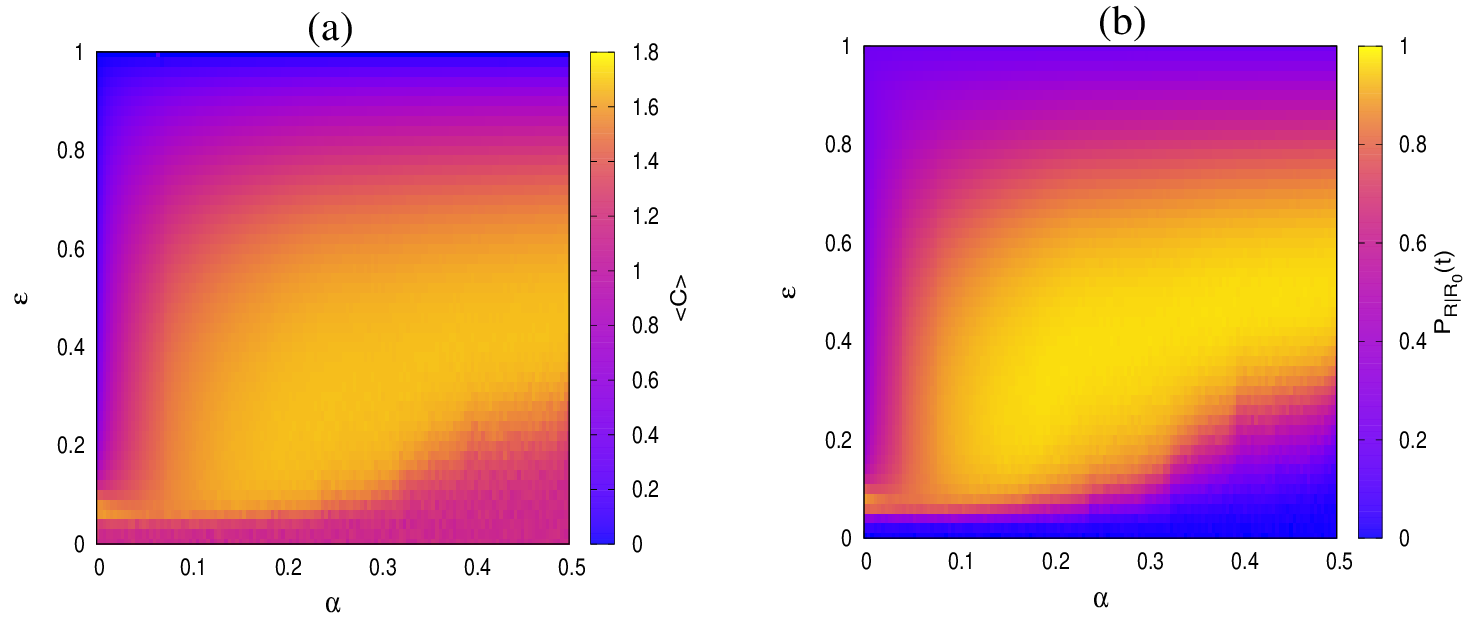}
\caption{(a) In the $\epsilon - \alpha$ plane, heat map of $\langle C \rangle$ after time $t=t_1$ starting from uniform initial condition. (b)
Heat map of ${\cal P}_{R|R_0} (t=t_1)$  in the same plane starting with non-uniform initial condition. We have used $t_1=10^8$ here. Both plots show an optimum region on $\epsilon - \alpha$ plane where RL algorithm is working best.} \label{fig:2sin_heatmap}
\end{figure}

In Fig. \ref{fig:2sin_heatmap} we show, in the $\epsilon-\alpha$ plane, the variation of uptake and  ${\cal P}_{R|R_0} (t)$, measured after a long time. We find both these quantities reach a maximum value for an optimal range of $\epsilon$ and $\alpha$. This shows that the RL strategy employed in this system works best in this range. 

\subsection{First passage time from lower peak to higher peak of attractant profile} \label{sec:fpt}

\begin{figure}[htbp!]
\includegraphics[scale=0.6]{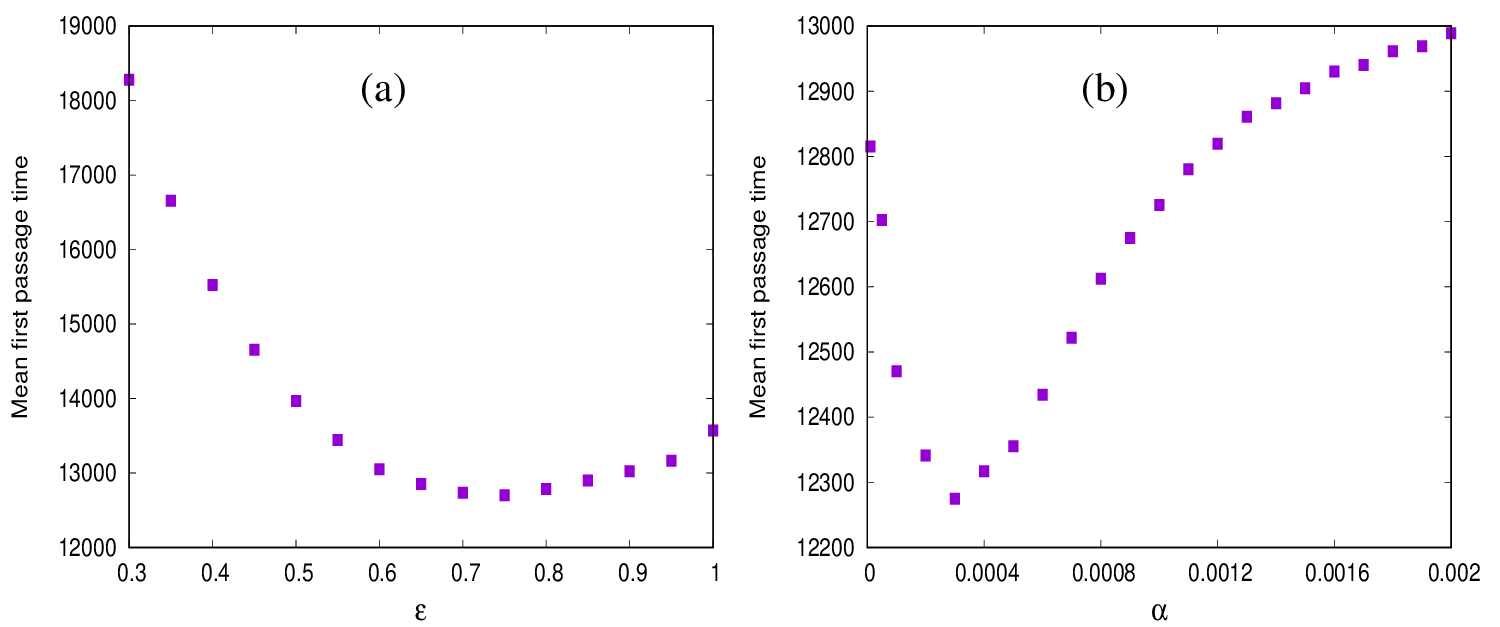}
\caption{Mean first passage time starting from the lower peak to the higher peak of the attractant profile. The quantity shows a minimum as a function of both $\epsilon$ and $\alpha$. For (a) $\alpha$ is fixed at $0.001$ and for (b) $\epsilon$ is fixed at $0.7$. All other simulation parameters are as in Fig. \ref{fig:px2sin}.}  \label{fig:fpt}
\end{figure}

To further probe the role of exploitation-exploration competition on the performance of the RL agent, we measure its mean first passage time from the lower attractant peak to the higher one. Starting from a local maximum, how long the agent needs to find the global maximum, is an important quantity to characterize its efficiency. For small $\epsilon$ the agent is not able to explore its surroundings and remains trapped near the local maximum. This increases its mean first passage time. On the other hand, for $\epsilon \to 1$ the agent shows little affinity towards attractant peaks. In this limit its motion is mainly diffusive which also corresponds to a large value of mean first passage time. For an intermediate $\epsilon$, mean first passage time shows a minimum (Fig. \ref{fig:fpt}(a)). For a given $\alpha$, there is an optimal value of $\epsilon$ for which the agent performs the quickest search. Fig. \ref{fig:fpt}(b) shows the plot of mean first passage time as a function of $\alpha$, for a fixed $\epsilon$. Even in this case, we find an optimum $\alpha$ corresponding to the quickest search. 

\section{Discussions} \label{sec:con}

In this work, we have considered an RL agent which is exploring its environment via run-and-tumble motion. We are interested in the question: under what condition the RL strategy is most efficient. We quantify efficiency by the probability to find the agent in the attractant-rich region in the long time limit, and also by how much the agent has learnt about its environment. A successful RL strategy allows the agent to quickly learn about its attractant environment and localize in the favorable region. We find depending on the nature of the attractant concentration profile, different RL strategies work best.

In the case when all the concentration peaks are of the same size, then starting from a uniform initial position, the agent is able to localize most strongly near the peak regions when exploration rate is low and learning rate is high. However, when the agent starts from the vicinity of one particular attractant peak, then it remains trapped there and even after a large time has passed, the agent is not able to learn about its complete environment if the exploration rate is low or learning rate is high. In this case, an optimum range of these two rates works best for the agent.

When the attractant profile has peaks of different sizes, the exploration-exploitation trade off is even more pronounced. Even when the initial position of the agent is uniformly distributed throughout the system, the agent is not able to learn about its complete environment and gets trapped in nearest attractant peak, for low exploration rate and high learning rate. As a result, even after a long time has passed, the agent may show same affinity for the lower and higher attractant peaks, which is detrimental for its performance. An optimum balance between exploitation and exploration helps the agent overcome the trapping effect and show its best performance. For non-uniform initial condition, when the agent starts near the lower attractant peak, its probability to localize near the higher peak after a long time, shows a maximum for intermediate rates of exploration and learning.

Our simple model has few limitations. For example, there is an important difference between our cost calculation method and the decision making process that an {\sl E.coli} cell follows during chemotaxis. As explained in Sec. \ref{sec:model}, our RL algorithm just takes into account the sign of the concentration difference experienced in the recent and distant past and does not care about the magnitude of the difference. But an {\sl E.coli} cell modulates its tumbling rate by taking into account both the sign and magnitude of the difference in concentration experienced in recent and distant past \cite{block1982impulse}. Most of our results presented here are for the case when the agent uses attractant levels at last two time-steps to read off the local gradient. This is the best choice to read off the sign of the local attractant gradient accurately. However, most of our qualitative conclusions remain valid (data not shown) even for larger values of $\Delta_1$ and $\Delta_2$, as long as they do not exceed typical run duration. For very large $\Delta_1$ and $\Delta_2$, the memory goes too far back in the past. The gradient calculated from $[L](x_1) - [L](x_2)$ may not accurately represent the current environment since the agent may have reversed its direction once or multiple times meanwhile.

 Throughout this work, we have considered one spatial dimension. Many experiments measure run-tumble motion of {\sl E.coli} in narrow microfluidic channel whose width is comparable to or smaller than typical run length \cite{PhysRevE.96.032409, PhysRevE.98.052413, KALININ20092439, jiang2010quantitative}, and therefore, the motion effectively takes place in one dimension. However, it is an interesting question to ask whether in a more natural setting, where the cell moves in two or three dimensions, our RL description might work. In two dimensions, for example, after each tumble the direction of the cell is randomized. Therefore, the effect of a tumble can not be simply described as velocity reversal, as we have done in the present model. In this case, one can classify the actions of the RL agent as `persist' and `randomize', where the former means persisting in the same direction, and the latter means choosing a random direction. During exploration, the agent can either persist or randomize with equal probability. Note that the possibility of randomizing direction after each tumble means that the trapping effect seen in Figs \ref{fig:p2p} and \ref{fig:2peak} is expected to be weaker in two dimensions. Indeed our preliminary data (not shown here) indicate that even when the exploration parameter $\epsilon$ is significantly small, the agent is able to escape a local attractant peak and explore its full environment. This affects the exploration-exploitation competition and exploration starts winning over exploitation at a much lower $\epsilon$ value, compared to what we have seen in one dimension. However, we verify (data not shown) that apart from some quantitative differences, all our qualitative conclusions remain valid in higher dimension as well.

Possible future directions include generalizing our simple model where state description can have a finer resolution (instead of only two possible values, high and low), or cost function can be assigned real values instead of $0$ and $1$, or using a learning parameter which evolves with time, etc. Also, the current study is limited for a single particle in a time-independent environment, the study can be extended for many interacting particles moving in an environment that changes with time.  Expanding the model for the case where particles also learn from each other might be interesting to explore in future studies.

\section{Acknowledgements}
RP acknowledges the research fellowship [Grant No. 09/0575(13013)/2022-EMR-I] from the Council of Scientific and Industrial Research (CSIR), India. SM thanks DST-SERB India, ECR/2017/000659, CRG/2021/006945 and MTR/2021/000438 for financial support. SM also thanks S.N. Bose National Centre for Basic Sciences, Kolkata for providing kind hospitality. SC acknowledges support from Anusandhan National Research Foundation (ANRF), India (Grant No: CRG/2023/000159).


\end{document}